\documentclass[review]{elsarticle}

\usepackage{amsmath,amssymb}
\usepackage{graphicx}
\usepackage{caption}
\usepackage{lipsum}
\usepackage{subcaption}

\usepackage[capposition=top]{floatrow}
\usepackage{pdflscape}
\usepackage{afterpage}
\usepackage{capt-of}
\usepackage{rotating}
\usepackage{newtxtext}
\usepackage{newtxmath}
\usepackage{tablefootnote}
\usepackage{bbm}
\usepackage{datetime}
\usepackage{multirow,tabularx}
\makeatletter
\def\ps@pprintTitle{%
 \let\@oddhead\@empty
 \let\@evenhead\@empty
 \def\@oddfoot{\centerline{\thepage}}%
 \let\@evenfoot\@oddfoot}
\makeatother

\biboptions{authoryear}

\begin{document}

\begin{frontmatter}

\title{
 Politicians' Willingness to Agree: \\ 
 Evidence from the interactions in Twitter of Chilean Deputies}
  

\author[rvt]{Pablo A. Henr\'iquez}
\ead{pablo.henriquez@udp.cl}
\author[rvt]{Jorge Sabat\corref{cor1}}
\ead{jorge.sabat@udp.cl}
\author[rvt2]{Jos\'e Patricio Sullivan}
\ead{jpsullivan@wustl.edu}

\address[rvt]{Faculty of Business and Economics, Universidad Diego Portales}
\address[rvt2]{Washington University in St. Louis}

\begin{abstract}
Measuring the number of ``likes" in Twitter and the number of bills voted in favor by the members of the Chilean Chambers of Deputies. We empirically study how signals of agreement in Twitter translates into cross-cutting voting during a high political polarization period of time. Our empirical analysis is guided by a spatial voting model that can help us to understand Twitter as a market of signals. Our model, which is standard for the public choice literature, introduces authenticity, an intrinsic factor that distort politicians' willigness to agree \citep{trilling2009sincerity}. As our main contribution, we document empirical evidence that "likes" between opponents are positively related to the number of bills voted by the same pair of politicians in Congress, even when we control by politicians' time-invariant characteristics, coalition affiliation and following links in Twitter. Our results shed light into several contingent topics, such as polarization and disagreement within the public sphere. 
\end{abstract}


\end{frontmatter}

\newpage
\section{Introduction}

Since \citet{poole1985spatial} dozens of papers have tried to infer politicians' ideology using voting data. Similar analysis have also been conducted using campaign contributions \citep{bonica2013ideology}, surveys and texts \citep{laver2014measuring}. \citet{bolstad2017categorization} suggest that these applications of the spatial voting theory can successfully identify clusters of political actors that can be associated to the same ideology (e.g. left and right-wing coalitions).\footnote{The empirical analysis of these political networks follows a revealed preference approach where politicians and voters' preferences could be characterised from their observed political behavior \citep{henry2013euclidean}.} However, this methodology is mostly agnostic of the many endogenous factors that determine the positions of political actors in the network, being very effective on localizing political actors but not on offering explanations for how and why they reached that position. In this paper we look into how political communication, understood as a channel of agreement and disagreement, can determine how politicians reach a given position within the political spectrum. 

We follow \citet{barbera2015birds}, who complements the analysis of politicians behavior using interactions in Twitter (e.g. ``retweets'' and ``following'' links) of U.S. political actors.\footnote{A ``retweet'' is a re-posting of a Tweet (message) from another user in Twitter. A ``follower'' is someone that decides to receive the Tweets of another user.} Our main contribution is to augment the analysis of politicians' behavior by combining their Twitter interactions with voting data from Congress. Specifically, we exploit Twitter ``likes'', a strong signal of appreciation from one user to another, to measure instances of agreement in the public political debate among politicians in Chile.\footnote{We understand the public political debate in a Habermasian sense, this is, as a public sphere where different actors try to advance their positions trough communicative rationality.} We use this empirical setting to test our main hypothesis, communications dynamic in Twitter does not translate into voting agreement, after controlling by by politicians' time-invariant characteristics, coalition affiliation and following links in Twitter.

Our main empirical result is that, after including the different controls, on average, politicians that agree in Twitter agree less in Congress. This empirically puzzling result is explained by a differential effect based on coalition affiliation. We find that, at the same time, the number of ``likes'' is negatively (positively) related to bills voted in Congress by coalition mates (opponents).

Our results can be interpreted using a public choice model. In the model, politicians use ``likes'' as signals for manipulating voters' beliefs. The model is standard in the sense that politicians decide based on electoral incentives (popularity). However, we include a non-standard mechanism based on authenticity. Following \citet{trilling2009sincerity}, we argue that politicians weight being consistent with their own ideology. We use authenticity as a factor that generates a trade-off between popularity and ideological consistency. Under the theoretical implications of our proposed public choice model, we suggest that our analyzed interactions are consistent with a fraction of politicians taking authenticity into account at the moment of deciding which ``tweets'' to ``like''. In addition, our model suggest that ``likes'' between opponents work as costly signals of political agreement, while ``likes'' between coalition mates would work as signals of group loyalty.

Chile is an interesting setting to test our hypothesis. Our interest is related to the different political cleavages that have shaped the political landscape since its return to democracy in 1990 \citep{bonilla2011social}, as well as, the evidence of ``centrifugal forces'' generated by the electoral system early documented by \citet{dow1998spatial}. After the return of democratic governments, Chile set up for a fast-paced economic development, sustained on free market-oriented policies and a stable political system, which transformed Chile in one of the best performing countries in the region \citep{barro1999determinants}. However, in October 2019, after almost three decades of sustained growth, the country’s social stability crumbled under huge protests and popular demonstrations against economic inequality. The political parties reacted by proposing a referendum to reform Pinochet’s Constitution. In October 2020, a referendum took place, with an overwhelming majority approving the rewriting of the constitution trough a democratically-elected Constitutional Assembly. Consequently, we argue that the current political situation in Chile offers a unique setting to analyze how politicians can reach agreement in a highly polarized political environment.

This paper intends to contribute to three strands of the literature. First, we contribute to the empirical spatial voting literature pioneered by \citet{poole1985spatial}, which focuses on identifying coalitions and other political associations using data on politicians' voting behavior, campaign contributions \citep{bonica2013ideology}, or surveys and text \citep{laver2014measuring}. Second, we contribute to the literature on disagreement, which acknowledges the importance of rationality as a mechanism to reach agreement \citep{habermas1998between,aumann1976agreeing}. This theoretical predictions based on rationality are inconsistent with persistent disagreement observed in societies ($cite$). The persistence of disagreement has been explained by behavioral theories of homophily \citep{halberstam2016homophily} or identity \citep{taylor2020agreeing}. Finally, our study has implications for research on polarization. \citep{zhang2008community} proposes modularity as a measure of polarization based on network theory. On the other hand, social scientists have studied polarization using a different approach, categorizing it as either ideological, affective or perceived \citep{fiorina2008political,iyengar2012affect,hetherington2016revisiting}.

The paper proceeds as follows. In Section \ref{sec:inst_fram} we describe the institutional framework. In Section \ref{sec:rel_lit} we review relevant related literature in spatial voting theory, political disagreement and polarization, highlighting how they connect with the present study. Section \ref{sec:data} provides a description of our three data sets: the CEP survey, interactions of politicians in Twitter and the bills voted in the Chilean Chamber of Deputies. Section \ref{sec:Model} outlines our public choice theoretical model. Section \ref{sec:emp_fram} we detail our empirical approach to test our proposed hypothesis. In Section \ref{sec:results} we present our results. Finally, in Section \ref{sec:con} we provide our main conclusions with some open ended discussion questions for future research.

\section{Institutional Framework}\label{sec:inst_fram}
Chile returns to democracy in 1989 after Pinochet's dictatorship. The Constitution established that Congress is divided in two Chambers. The Chamber of Deputies with 120 members, and the Senate with 38 members. The electoral system that ruled the allocation of seats in both chambers was a variant of the well studied D'Hondt electoral system, a system locally known as ``Binominal''. As has been analyzed by \citet{dow1998spatial}, the Chilean variant forced the formation of two coalitions formed by multiple parties. This incentive is set by mandating that each coalition presents two candidates by district. The ``Binominal'' system awarded the first seat based on the largest number of votes by coallition. However, if the combined number of votes is more than 2/3 of the total votes in that district, the candidates of the same coalition win both seats. Otherwise, one seat goes to each coalition. Authors have suggested that proponents of the system believe that it maintained political stability, at the same time that caused \textit{malapportioned} \citep{riquelme2018voting}. Moreover, \citet{dow1998spatial} suggested that favor reward extremists and punish moderates. 

In 2009, after 37 years without parliamentary representation, the Communist Party returned to Congress, trough the election of 3 deputies. This electoral victory was made possible by a deal with the left wing coalition, which presented no candidate in some districts. Later, in 2013 the Communist Party and other independents group joined the left wing coalition for the Presidential and Congressional Elections. One of the main objectives of this coalition was to reform the electoral system to increase the representation of new groups that have been excluded from parliamentary representation. 

In 2017 the Chilean electoral system was reformed for a traditional D'Hondt system. This reform has coincided with the downfall for the traditional left, and the increasing representation of the radical left and right wing groups \citep{bunker2020electoral}. Moreover, \citet{fabrega2018polarization} suggest that this electoral reform is related to a latent political polarization that started in the 2000s.\footnote{\citet{fabrega2018polarization} suggest that the student protests of 2006 and 2011 are early signs of their measured political polarization.} 

In October 2019, after almost three decades of sustained growth, Chile experienced a series of protests, an political unrest against economic inequality. The political parties reacted by proposing a referendum to reform Pinochet’s Constitution. In October 2020, a referendum took place, with an overwhelming majority approving the rewriting of the constitution trough a democratically-elected Constitutional Assembly.

During 2020 Chile has been affected by the COVID crisis, which in the context of the 2019 social outbreak, has been a new source of disagreement for Chilean politics \citep{oyarzun2020chile}. 

\section{Related literature}\label{sec:rel_lit}
Our contribution to the literature is threefold. First, we contribute to the empirical spatial voting literature that focus on recovering estimates of politicians' preferences (e.g. ideological positions) from their observed voting behavior. A methodological approach pioneered by \citet{poole1985spatial}. In more recent times, the same approach has included other types of data sets, such as campaign contributions \citep{bonica2013ideology}, or surveys and texts \citep{laver2014measuring}. As \citet{bolstad2017categorization} suggest, applying this methodological approach, several studies can successfully claim that they are able to identify coalitions and other political associations. However, this literature tend to be agnostic of the factors that can explain why two politicians are voting in favor of the same law, other than the estimated ideological closeness. In this paper, following \citet{barbera2015birds} we augment voting data with interactions from Twitter. As opposed to \citet{barbera2015birds}, who analyzes 'retweets', we focus on 'likes' among deputies. A more direct signal of appreciation that has been used to understand social networks in other contexts \citep{wachs2017men}.

Second, we contribute to the literature that studies political disagreement. \citet{hinich1996ideology} propose that ideologies are a way to resolve political disagreement with respect to three questions: what is good, who gets what and who rules. In this interpretation, ideologies work as theorizations of politicians' own position in the political arena \citep{martin2015ideology}.\footnote{\citet{martin2015ideology} proposes that ideologies are not necessarily internally consistent. However, ideologies would provide a representation of political alliances, and more importantly, the nature of political opponents.} In practice, political disagreement would take the form of arguments.\footnote{\citet{hinich1996ideology} cites the case of the minimum wage debate analyzed by \citet{friedman1966essays}. Political disagreement in this case can be easily disentangled into values, and different predictions over the effects on the poverty rate or the labor market.} In the ``ideal speech situation'' posed by \citet{habermas1979universal}, if all parties seek the truth and they do not behave strategically, this would lead to the ``best argument'' to win, as has been argued by \citet{habermas1998between}. Moreover, in theory, is well known since \citet{aumann1976agreeing} that, if it is common knowledge that both parties are Bayesian rational, they cannot agree to disagree on matters of fact. These theoretical results enter in contradiction with the empirical evidence that suggests that disagreement is a very persistent process. A finding that has also been associated with behavioral theories of homophily \citep{halberstam2016homophily} or identity \citep{taylor2020agreeing}. At the same time, other authors argue that the internet have exacerbated the emergence of echo chambers \citep{sunstein2008neither}. We contribute to this literature combining new data from Twitter, that allows us to identify instances of agreement between politicians, and matching it with bills voted in favor in Congress. Trough a political economy model, we argue that our analysis can shed light on the willingness to agree of politicians. 

Third, we contribute to the broad literature on political polarization. \citet{zhang2008community} proposes to study polarization using network theory, where the network's modularity is calculated using votes by U.S. Congress members.\footnote{Intuitively, the modularity corresponds to an aggregated measure of the intra-coallition versus inter-coallition interactions. For more details see \ref{sec:Community_detection}.} On the other hand, social scientists have studied polarization using a multidisciplinary approach, understanding the phenomena either as ideological, perceived or affective. Ideological polarization measures the differences among policy choices and political positions. For example,  \citet{fiorina2008political} argued that there is no conclusive evidence of higher polarization of citizens' ideological positions in the U.S.
Perceived polarization accounts for how individuals perceive more polarization than actually exists. Thus assuming that out-party
members are farther away from them on the ideological continuum than they
actually are, according to measures of their survey preferences \citep{enders2019differential}. Affective polarization measures the animosity between political opponents, heavily dividing the spectrum between in-group and out-group. \citet{iyengar2019origins} found that Democrats and Republicans both say that the other party's members are hypocritical, selfish, and closed-minded, and they are unwilling to socialize across party lines.\footnote{\citet{westfall2015perceiving} provides evidence of higher perceived polarization among individuals that identify more with the Republican or Democratic parties in the U.S.} Our paper intends to contribute to the multidisciplinary literature on polarization trough the empirical analysis of a polarized society (measured by modularity) during the 2019-2020 period. A society that has experienced increasing levels of polarization in the last two decades according to \citet{fabrega2018polarization}. In contrast to most of polarization studies, we focus on the instances of depolarization, analyzing cross-cutting votes and ``likes''.

\section{Data}\label{sec:data}
\subsection{Data sources}
In this paper we combine different public data sources, from a public opinion survey conducted by the Centro de Estudios Públicos (CEP), Twitter, and the Chamber of Deputies of Chile.
\subsubsection{CEP political opinion survey}
We study Chile's ideology distribution of voters from the CEP political opinion survey because it intends to be nationally representative, and it has been published periodically since the 90s. We mainly focus in the May 2019 version.\footnote{The field work last 2 months approximately.}\footnote{This data has been used to analyze Chilean politics in the empirical Public Choice literature, see \citet{bonilla2011social}.} We choose this version because it precedes Chile's political unrest event of October 2019. The study surveyed 1,380 citizens of the 128 most important counties in Chile. In addition to different socio-demographic questions, the study also ask about political opinions of well-recognized politicians, and their self-reported ideology, in a discrete 1 to 5 scale. In Figure \ref{fig:CEP_distribution}, we highlight how the estimated non-parametric distribution is far from a normal distribution, where most of voters would be bunched around the center. It is worth mentioning that Figure \ref{fig:CEP_distribution} considers the roughly 55\% of people that do not self-identify with any of these categories, a common phenomena in Latin American politics \citep{ruth2016clientelism}.\footnote{In our calibration of the model, unknown voters are going to be  uniformly distributed in the 1 to 5 axis.} From the same opinion survey, we can recover politicians ideology using the self-reported ideology variable described above, and the public opinion of well-known politicians.\footnote{Our proposed method is explained in \ref{sec:OLS_estimate}.} In Figure \ref{fig:Politician_Estimates2}, we plot the regression coefficients that measure politicians' ideology. Interestingly, our estimates suggest that voters would not see a lack of supply of politicians with different ideologies.\footnote{For a more detailed analysis of the sampled 28 politicians see Figure \ref{fig:Politician_Estimates1}.}

\subsubsection{Twitter}
Twitter ``likes'' are obtained using Twitter API. We generate a script to collect the last 500 ``likes'' for every Chilean deputy that can be tracked in Twitter. Data extraction occurs at different points of time during the 2019-2020 period, which covers October 2019 Chile's political unrest, and the first year of the COVID crisis.\footnote{For a description of time periods see Table \ref{tab:DatesData}.} The Twitter database is structured as a panel dataset, where we can track the number of ``likes'' between deputies during the studied time window. In addition, from the same source we can record if one politician follows another. On average, politicians in our sample send 2.74 messages in Twitter (``tweet'')  a day. However, the number of messages per day by politician is highly skewed, with the top ten users ``tweeting'' more than 8 times a day. In Table $\ref{tab:sum_stat}$ we can see that the number of interactions that we are following in Twitter is increasing over time. This is explained by the higher number of deputies start using Twitter after the Chilean unrest event of October 2019, and also after the COVID crisis started. The average number of ``likes'' is 0.45 for the whole period. However, this number can vary over time as is observed in Table $\ref{tab:sum_stat}$. From the same table, we can see that the probability of observing a deputy giving a ``like'' to an opponent (Opponents Mean) is less than a tenth of the pooled mean (Opponents All). On the other hand, we also find that the unconditional probability that a deputy follows an intra-coalition mate is roughly 47\%, which compares higher with the approximately 18\% probability that a politician follows an opponent.

\subsubsection{Chilean Chamber of Deputies}
Voting records are obtained from the website of the Chilean Chamber of Deputies. Using data scrapping, we collected all voting records by politician, before Twitter data is collected. We collected bills voted around 40 days before interactions in Twitter are retrieved. We do this as an intent to study the interactions in Twitter and Congress during the same time window. In Table $\ref{tab:sum_stat}$ we can see that, on average, the number of bills voted in favor among all deputies is not very different to the number of bills voted between opponents. This finding is consistent with politicians disagreeing more in Twitter than in Congress.

\begin{table}[H]
  \centering
  \caption{Summary Statistics}
    \begin{tabular}{clccccc}
          & \textbf{Tweet likes} & \textbf{May-19} & \textbf{Oct-19} & \textbf{Feb-20} & \textbf{Jun-20} & \textbf{Dec-20} \\
    \multirow{4}[0]{*}{All} & Mean  & 0.47  & 0.48  & 0.41  & 0.39  & 0.49 \\
          & Median & 0.00  & 0.00  & 0.00  & 0.00  & 0.00 \\
          & Std. Dev. & 5.80  & 5.84  & 5.12  & 4.50  & 5.49 \\
          & N     & 8113  & 8640  & 8640  & 10530 & 12727 \\
    \multirow{4}[0]{*}{Opponents} & Mean  & 0.02  & 0.02  & 0.04  & 0.03  & 0.05 \\
          & Median & 0.00  & 0.00  & 0.00  & 0.00  & 0.00 \\
          & Std. Dev. & 0.21  & 0.21  & 0.45  & 0.29  & 0.48 \\
          & N     & 3980  & 4252  & 4252  & 5180  & 6339 \\
          & \textbf{Votes in favor} & \textbf{May-19} & \textbf{Oct-19} & \textbf{Feb-20} & \textbf{Jun-20} & \textbf{Dec-20} \\
    \multirow{4}[0]{*}{All} & Mean  & 77.02 & 91.78 & 82.52 & 72.93 & 82.33 \\
          & Median & 77.00 & 92.00 & 76.00 & 74.00 & 75.00 \\
          & Std. Dev. & 23.72 & 28.25 & 35.33 & 17.19 & 41.36 \\
          & N     & 8113  & 8640  & 8640  & 10530 & 12727 \\
    \multirow{4}[0]{*}{Opponents} & Mean  & 68.64 & 91.96 & 85.30 & 73.61 & 81.29 \\
          & Median & 68.00 & 92.00 & 78.00 & 75.00 & 77.00 \\
          & Std. Dev. & 19.32 & 27.44 & 37.59 & 16.20 & 32.15 \\
          & N     & 3980  & 4252  & 4252  & 5180  & 24003 \\
    \end{tabular}%
  \label{tab:sum_stat}%
\end{table}%

\subsection{Network construction}\label{sec:networks_methodology}

\begin{figure}[H]
\centering
\includegraphics[width=\textwidth]{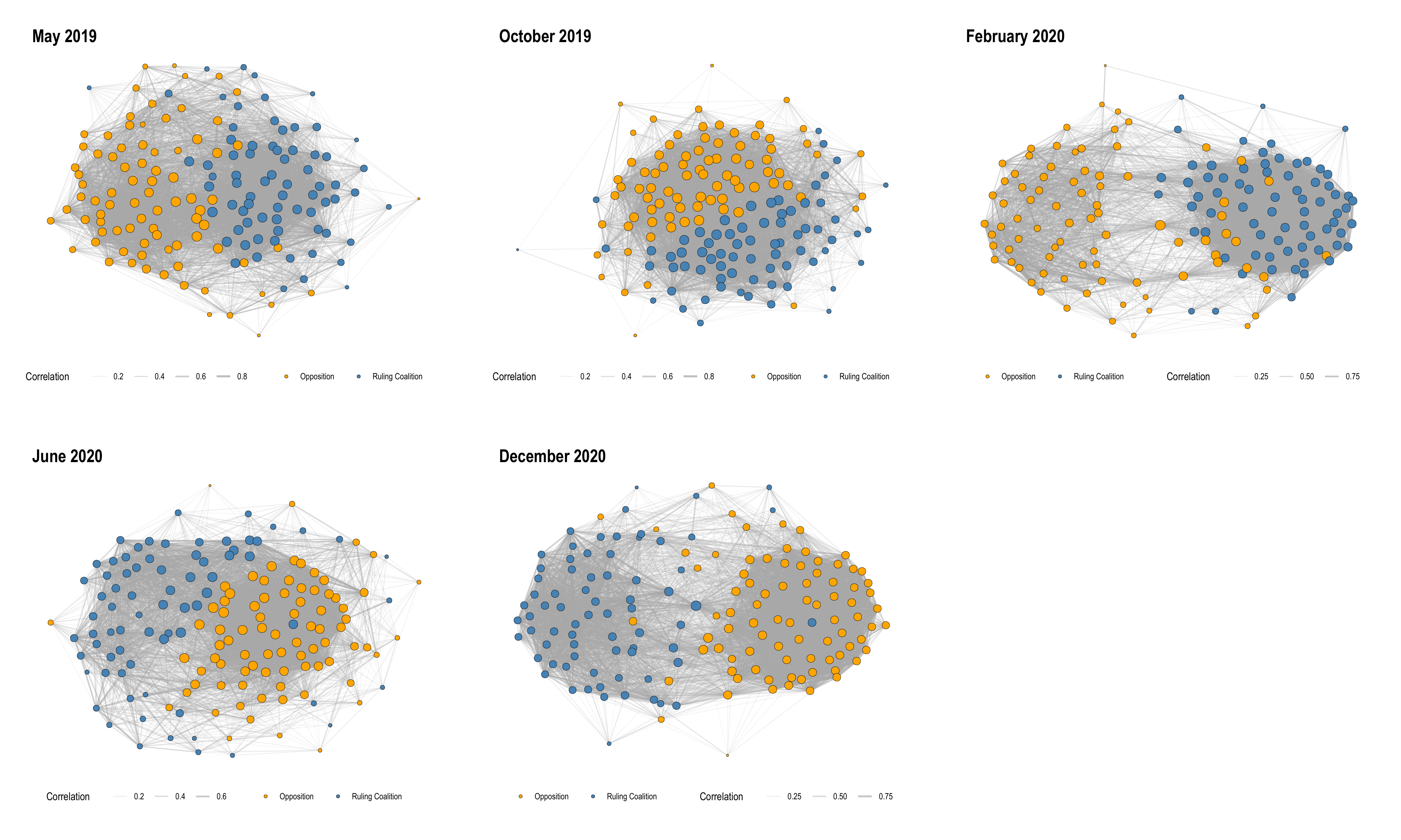}
\caption{Network analysis of bills voted in favor.}
\label{network_votes}
\end{figure}

\begin{figure}[H]
\centering
\includegraphics[width=\textwidth]{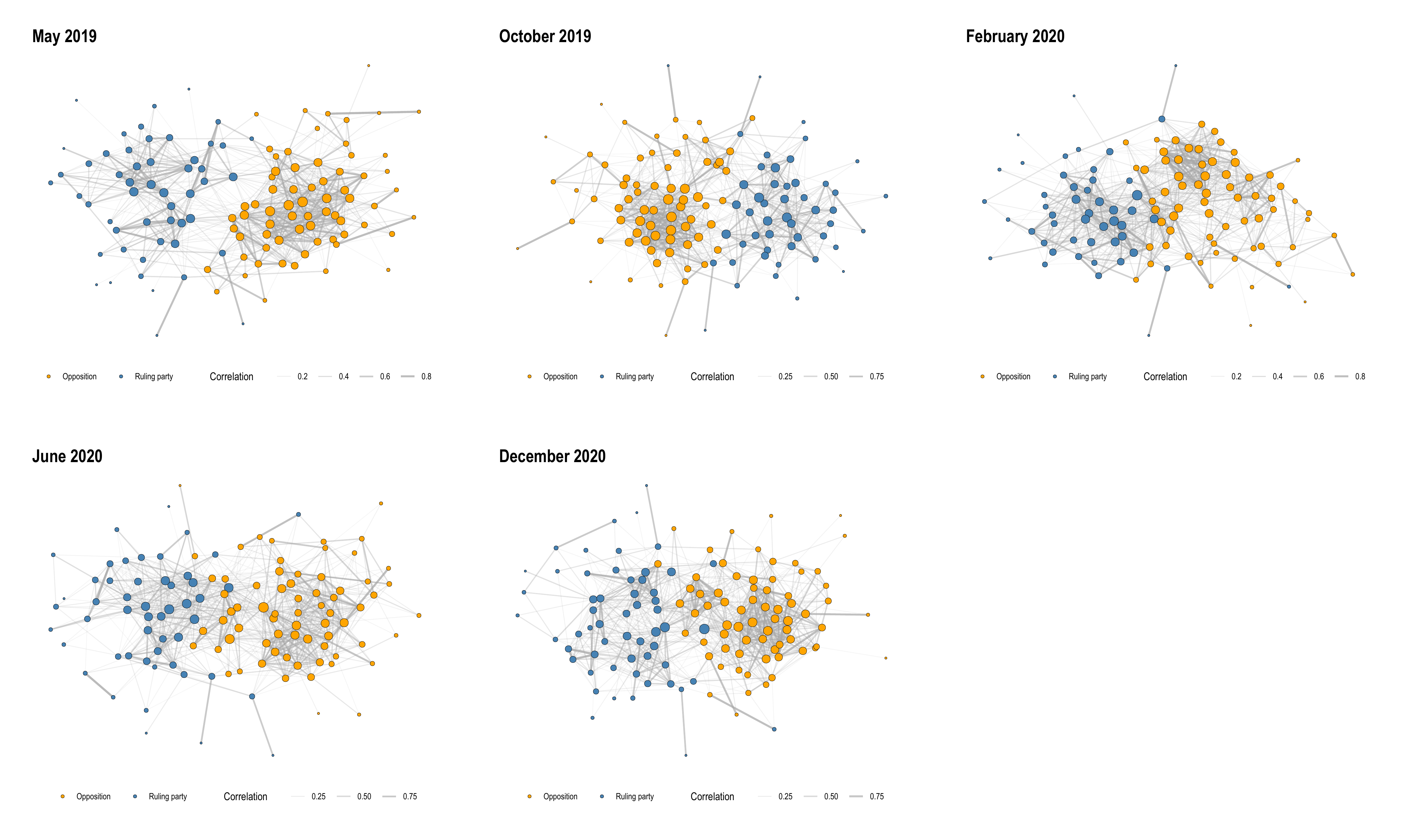}
\caption{Network analysis of ``likes''.}
\label{network_likes}
\end{figure}

In this section we describe the construction of indirect networks based on the interactions of politicians in Twitter and Congress. Specifically, we construct weighted networks $G_{i}=(V_{i},E_{i},W_i)$, where $V_{i}$ is the set of nodes corresponding to ``likes'' or votes in favor between politicians, $E_{i}$ is the set of edges between them, and edge weights ($W_i$) corresponds to the Pearson correlation coefficient. By setting a threshold $\theta=0.1$, we can transform correlation matrices into correlation networks. In Figures \ref{network_votes} and \ref{network_likes} the networks of votes in favor and ``likes'' are presented. As we can see, consistently with the summary statistics presented in Table \ref{tab:sum_stat}, the density of ``likes'' between politicians is lower than in Congress. In other words, agreement between deputies appear to be more common in Congress than in Twitter. On the other hand, the network figures also suggest that deputies tend to interact more with politicians from the same coalition, which is consistent with the evidence of homophily documented by \citet{halberstam2016homophily} using Twitter data related to politically-engaged users.

\begin{figure}[H]
\centering
\includegraphics[width=\textwidth]{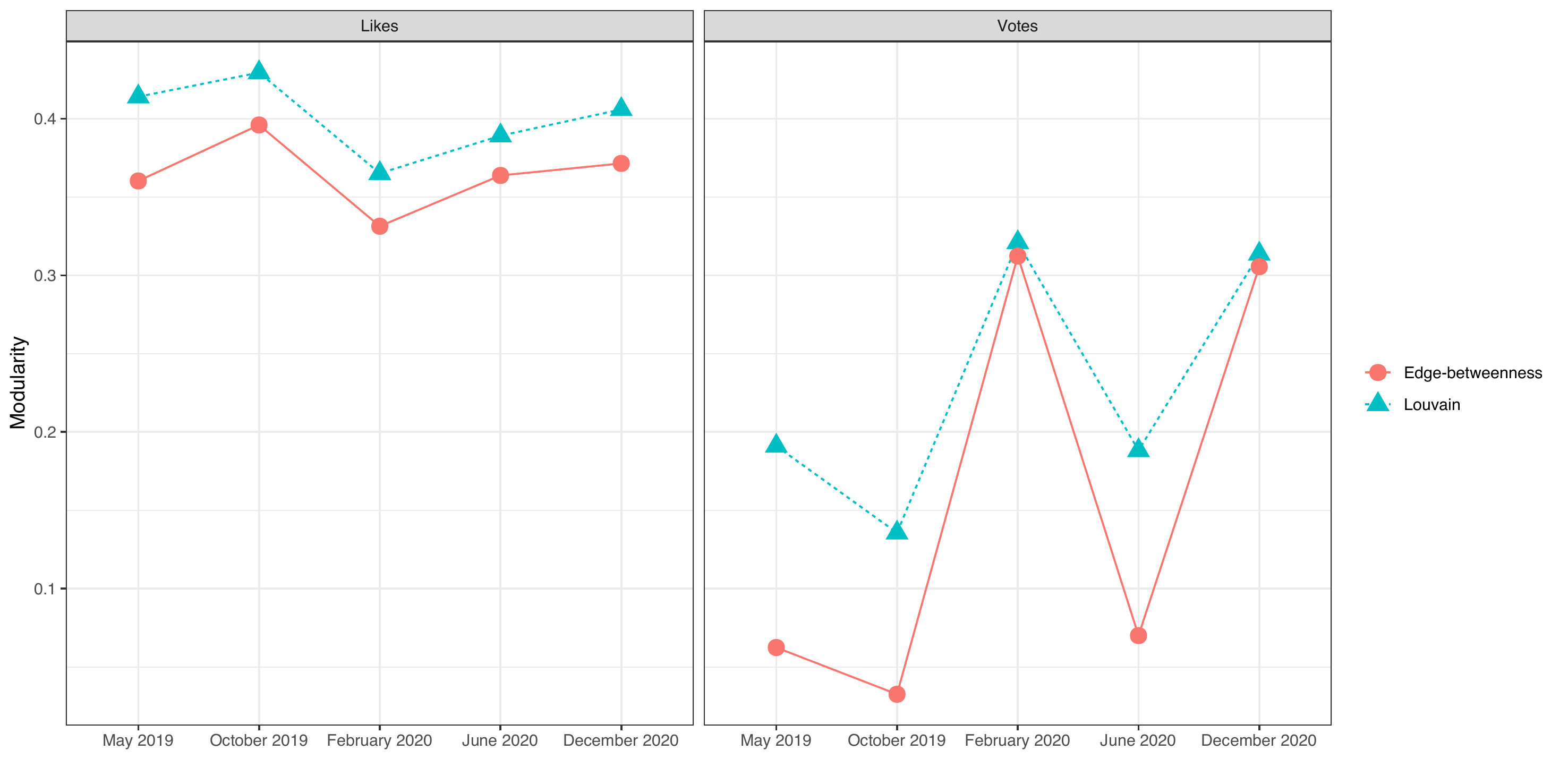}
\caption{Evolution of the modularity measurement. We considered the groups obtained via community detection method.}
\label{modularity}
\end{figure}

In order to provide a further characterization of the networks we follow \citep{zhang2008community}, which proposes modularity as a measure of political polarization. We calculate the modularity of the network following \citet{Newman2006} \footnote{See \ref{sec:Community_detection} for more details.}. In our application of his method we need to first identify relevant communities in Congress and Twitter to then quantify the severity of such a divisions. Intuitively, modularity measures network segregation into these distinct identified communities. For example, a network with high modularity would be divided into clusters having many internal connections among nodes, and few connections to other communities. In the literature, we find different ways to identify communities \citep{ribeiro2018dynamical}. In our paper we use the Louvain algorithm \citep{blondel2008fast}, as well as the Edge betweenness algorithm \citep{newman2004finding} as a robustness check.\footnote{The Louvain method is based on a greedy optimization method, while the edge betweeness algorithm is an iterative process based on betweenness centrality.} In Figure \ref{modularity} we present the evolution of the modularity of the networks constructed with data from Congress and Twitter. Our quantitative analysis confirm the intuition provided by Figure \ref{network_votes} and Figure \ref{network_likes}. First, estimated polarization is notably higher in Twitter than in Congress, until the Chilean political unrest event of October 2019. However, our results suggest that this event, as well as the COVID crisis are associated with an important increase in the polarization in Congress.\footnote{The decline in modularity of the network of votes in favor by June 2020 is related to the cross-party agreement of an emergency coronavirus plan.}\footnote{Our results appear to be robust to the methodological choice of community detection algorithm.} Finally, is important to highlight that our measured levels of polarization in Twitter ($\approx0.4$) are consistent with the highest levels of modularity found by \citet{waugh2009party} using roll call votes from the U.S. House during the 1788-2004 period, and \citet{conover2011political} findings obtained from a network of ``retweets'' related to politically relevant ``hashtags'' during the congressional midterm election of 2010.

\section{Theoretical model}\label{sec:Model}
In this section, we outline a public choice model that is used to interpret the cross-section of ``likes'' in Twitter among Congress members. The model considers a finite number of politicians that compete in a voting market. We assume that the voting market is populated by politicians that have different ideology types. An ideology type $i$ is well described by draws from a normal distribution with parameters $\mu_i$ and $\sigma_i$.\footnote{The practical implication of this assumption is that, in our model, some politicians are, exogenously more ideologically flexible than others, regardless of their ideological positions.} On the other hand, voters form beliefs about politicians' ideological positions. Voters can be simply characterised by a discrete distribution in a one dimension political axis (right to left wing). Society is fragmented into K political groups with ideology ($i_k$) and a population weight ($p_k$).\footnote{The specific parameter assumptions that govern the electoral population are shown \ref{sec:MedianVoter}}. Specifically, we highlight that in our model a more normally distributed electoral population is related to less polarization among elected politicians.

The dynamic of the political competition in the model occurs within coalitions, or ideology groups. This assumption is related to the D'Hondt proportional representation rule that governs the Chilean Chamber of Deputies, which seeks to elect Congress members in proportions that better represent the electoral population.\footnote{Until 2017 Chile had a modified D'Hondt system with two seats by electoral entity \citep{dow2001comparative}. It has been argued by this system forced bipartisanship trough the formation of two coalitions in order to maintain political stability \citep{riquelme2018voting}.} In the model, $N$ politicians sort into $n$ coalitions. Then, within the coalition, politicians gain electoral power based on the minimum distance to all competitors in the group. Ideological distance between politician's $i$ with respect to group of voters $k$ is calculated as follows:
\begin{equation}
\begin{aligned}
d_{i,k}=\big(\mu_i-i_k\big)\dfrac{1}{w_k}
\end{aligned}
\end{equation}
\noindent where a positive (negative) value for $d_{i,k}$ measures how much to the left (right) is politician's $i$ from group $k$, and $\dfrac{1}{w_k}$ is a factor that penalizes the focus on smaller groups of voters.

Politicians, within a coalition, compete based on distance $d_{i,k}$. Political power within the coalition is related to votes, which are endogenously determined by the distance between politician $i$ and political group $k$: 
\begin{equation*}
    \mathbbm{1}_{i,k} = \begin{cases}
              1 & \text{if } d_{i,k}=min\big\{d_{i,1},...,d_{i,K}\big\},\\
               0 & \text{if } $\text{otherwise}$
          \end{cases}
\end{equation*}

By assumption, within the group, political power depends on the number of votes (weighted by population) that politician obtain:

\begin{equation}
\begin{aligned}
v_{i}=\sum_{k=1}^{K} \mathbbm{1}_{i,k} w_k
\end{aligned}
\end{equation}

The ideology of the political front-runner, within the coalition, is based on the highest number of votes within the coalition:

\begin{equation*}
    \mu^f_k = \begin{cases}
              \mu_{i} & \text{if}\;\; v_{i,k}=max\big\{v_{1},...,v_{N_k}\big\},\\
               0 & \text{if}\;\; $\text{otherwise}$
          \end{cases}
\end{equation*}
\noindent where $N_k$ is the number of politicians that compete in coalition $k$.

In our model, politicians use Twitter to manipulate voters' posterior beliefs (as in a Bayesian persuasion game?). Manipulation occurs trough the effect of a signal that a politician $i$ sends to voters at the moments of giving a ``like'' to a message sent by another politician $j$ ($\delta_{i,j}$). Under the assumption that signals come from a normally distributed likelihood function with known variance equal to $\sigma_i$. Posteriors beliefs from the voters perspective can be determined as simple weighted average:
\begin{equation}
\mu_i^{*}=\dfrac{\mu_i}{1+\omega}+\dfrac{\omega}{1+\omega} \delta_{i,j}
\end{equation}
\noindent where $\omega$ captures how much voters weight the signal ($\delta_{i,j}$) versus their prior view ($\mu_j$) on politician $j$.

We define the endogenous variable $l_{i,j}$ as a dummy variable that takes a 1, if and only if, politician $i$ gives a ``like'' to politician $j$'s message, and zero otherwise. Assuming self interest preferences, a politician decides to ``like'' a message of another politician $j$ if it increases politicians' popularity ($\Delta P(l_{i,j})$). Politician's popularity is measured by distance, or proximity, in a right-left axis as has been empirically suggested by \citet{busch2016estimating} using data from Europe.

On the other hand, we incorporate an additional factor that influences politicians' behavior motivated by the idea of authenticity developed by \citet{trilling2009sincerity}. We include an authenticity parameter ($\gamma$) that measures politician's capacity to reach agreement in public argumentation \citep{haberrmas2006political}. In other words, politicians apply a factor $\gamma$ to the disutility of giving a ``like''  to some political message that is far from its own ideology ($\mu_i$). In Section \ref{sec:model_simmulation} we show that his parameter generates ideological consistency. However, at the same time reinforces the propensity to give a like to a message that is ideologically appealing, regardless of the identity of the message sender. In our model, this ideological distance is measured by $\Delta A(l_{i,j})$. The decision of politician $i$ to give a ``like'' to politician $j$'s message ($\delta_{i,j}$) can be formulated as follows:
\begin{equation}
\begin{aligned}
\max_{l_{i,j}} \quad & \Delta P(l_{i,j}) - \gamma \Delta A(l_{i,j})\\
\end{aligned}
\end{equation}
 \noindent where $\Delta P(l_{i,j}=1)$ measures the absolute change of politician $i$'s ideological distance with her front-runner candidate ($f$), before ($|\mu_{i}-\mu^f_k|$) and after giving the ``like'' ($|\mu_i^{*}-\mu^f_k|$). A positive (negative) value for $\Delta P$ measures electoral gains (losses) out of giving a ``like'' to a message $\delta_{i,j}$. On the other hand, $\Delta A(l_{i,j}=1)$ measures how far is the message from politician's own ideology ($|\delta_{i,j}-\mu_{i}|/\sigma_{i}$). 
 
 \subsection{Model Parameters}\label{sec:model_param}
\begin{figure}[htbp]
  \includegraphics[width=\linewidth]{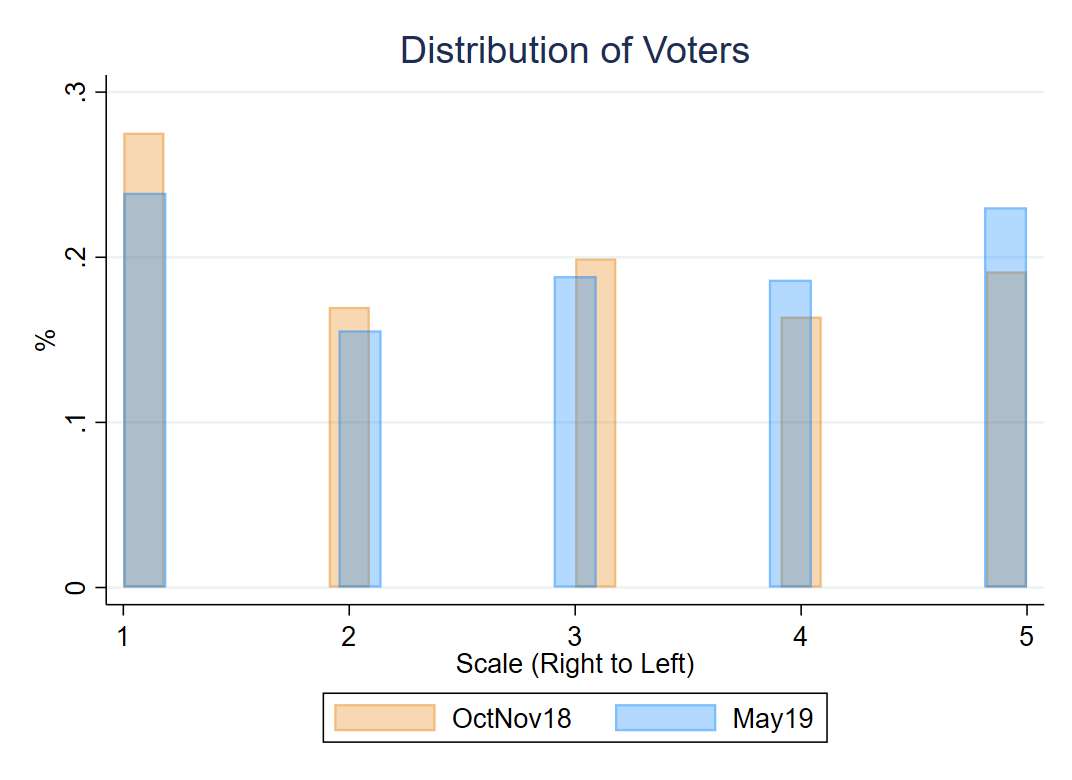}
  \caption{The figure measures the percentage of people in Chile that is associated with each ideology group, a scale from 1 to 5. Null responses are uniformally distributed.}
  \label{fig:CEP_distribution}
\end{figure}

We solve the model for a set of parameters that we took from the CEP opinion survey. In the simulations of the model, we use the ideological levels, as well as the proportion of voters documented in Figure \ref{fig:CEP_distribution}.\footnote{In \ref{sec:MedianVoter} we show that the assumed distribution of voters has implications that are related to violations of the median voter theorem. In our baseline case, some politicians have incentives to move toward the extremes of the political spectrum.}

\begin{figure}[H]
  \includegraphics[width=\linewidth]{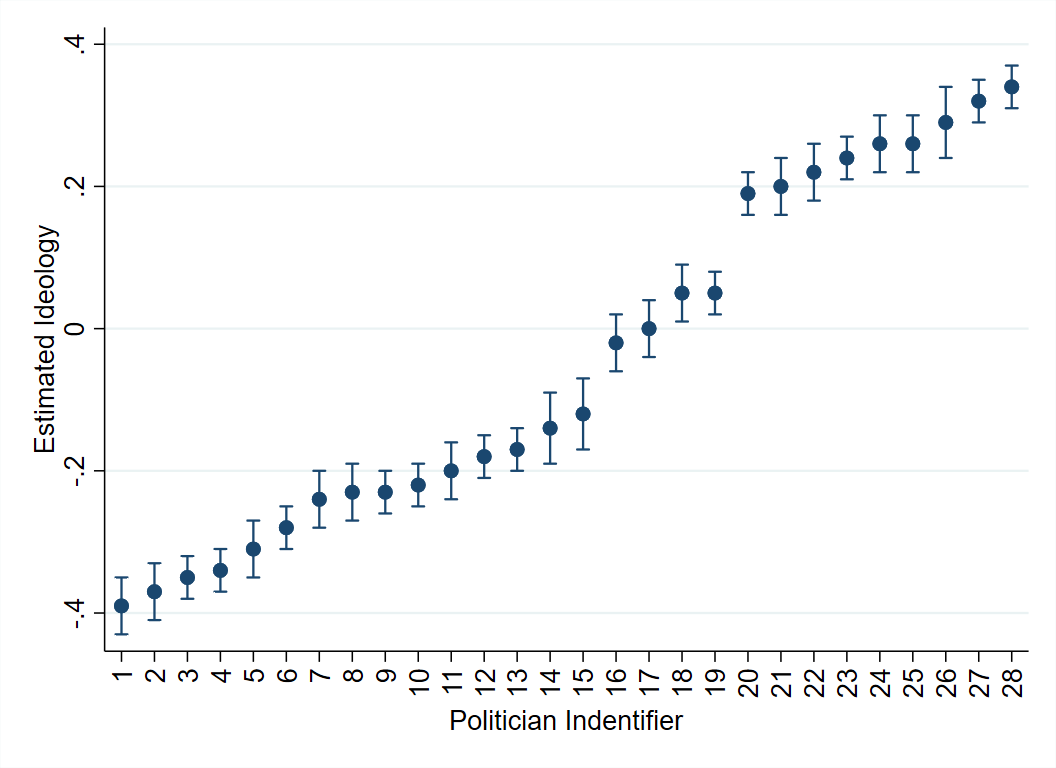}
  \caption{The figure plots the mean and the standard deviation of the ideological distribution of each politician in the model. Politicians are sorted from right to left.}
  \label{fig:Politician_Estimates1}
\end{figure}

Parameters that describe politicians' ideology ($\mu_i$ and $\sigma_i$) are also taken from the same opinion survey. Estimates are obtained by an OLS regression of surveyed subjective evaluation of a set of 28 politicians on self reported ideology. Our estimates allow us to have a measure of public opinion perceived ideology for a set of Chilean politicians. Intuitively, this measure is based on the correlation between subjective opinion and self reported ideology .\footnote{For a detailed description of our estimation see \ref{sec:OLS_estimate}} The mean and standard error of the coefficients plotted in Figure \ref{fig:Politician_Estimates1} determine politicians' ideology ($\mu_i$ and $\sigma_i$).

\newpage

 \subsection{Model Simulation}\label{sec:model_simmulation}
In this section, we present the predictions of our proposed model. It is important to highlight first that our results are based on $\omega=1$, which is equivalent to assuming that voters equally weight politicians' prior ideology and the observed signal related to the given ``like''. Second, in the model the political market (e.g. politician and voters' ideology distribution) is well described by the parameters presented Section \ref{sec:model_param}.

\begin{figure}[H]
    \centering
    \subfloat[]{\scalebox{0.25}{\includegraphics{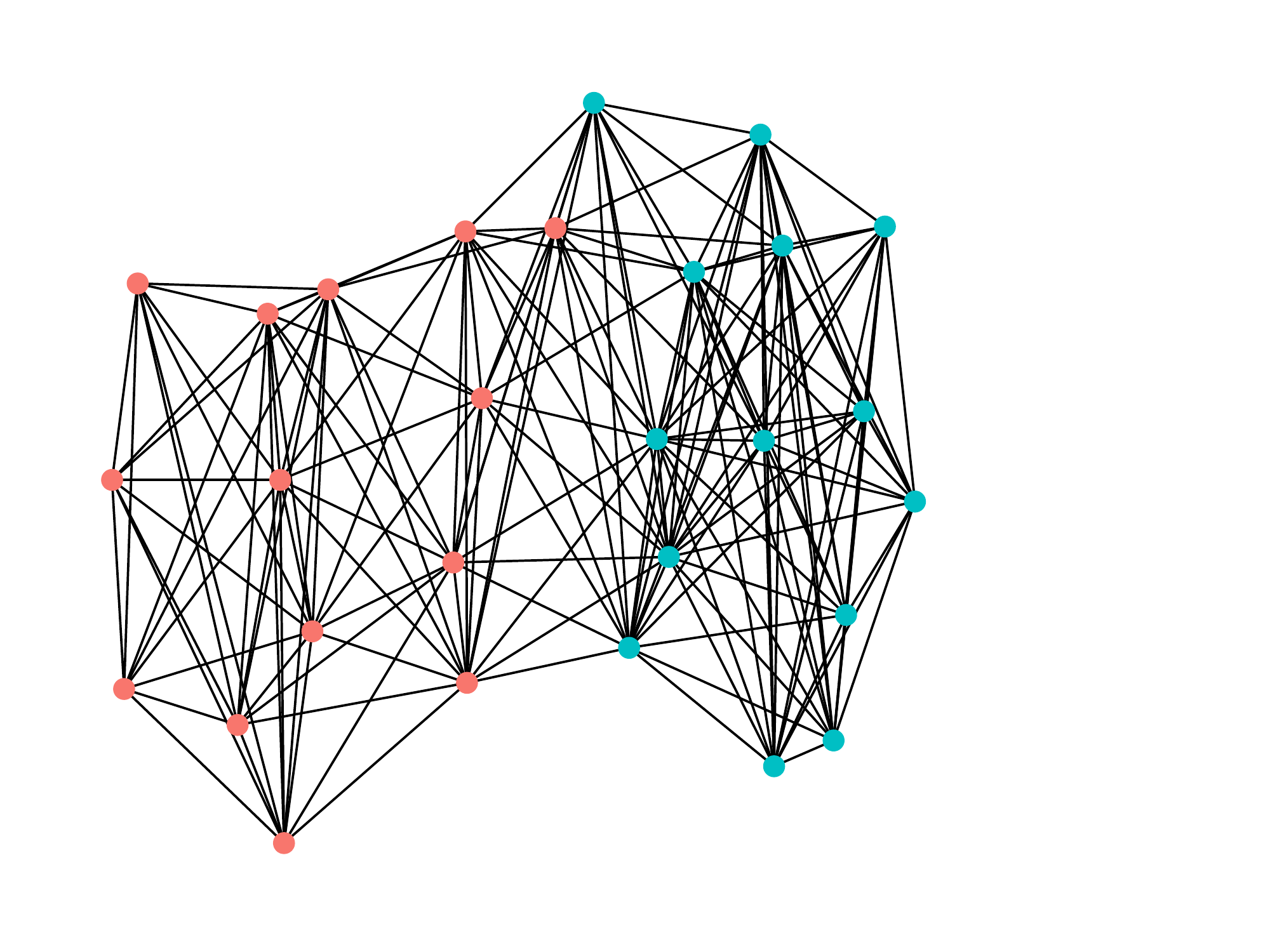}}}
    \hfill
    \subfloat[]{\scalebox{0.25}{\includegraphics{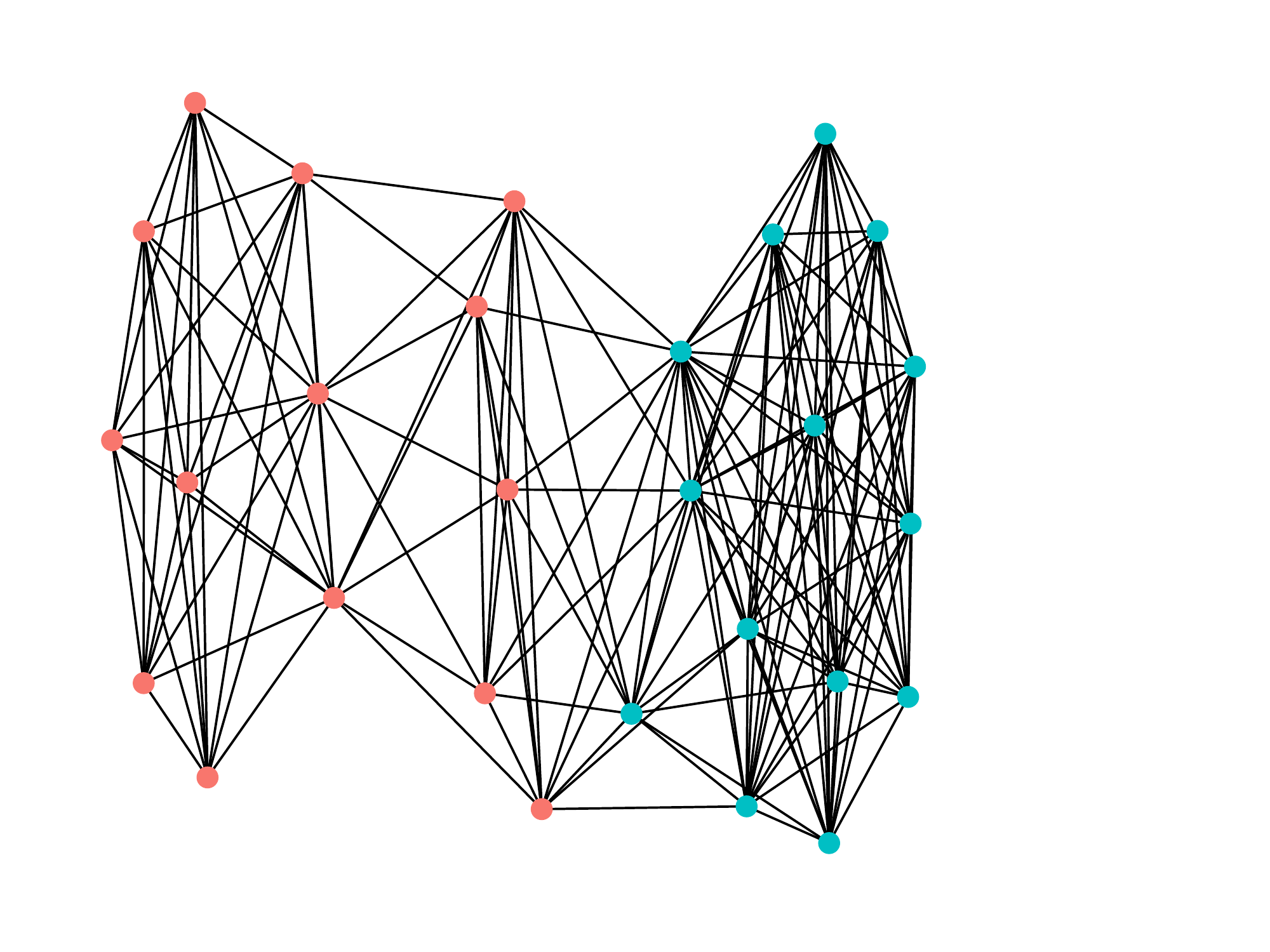}}}
    \hfill
    \subfloat[]{\scalebox{0.25}{\includegraphics{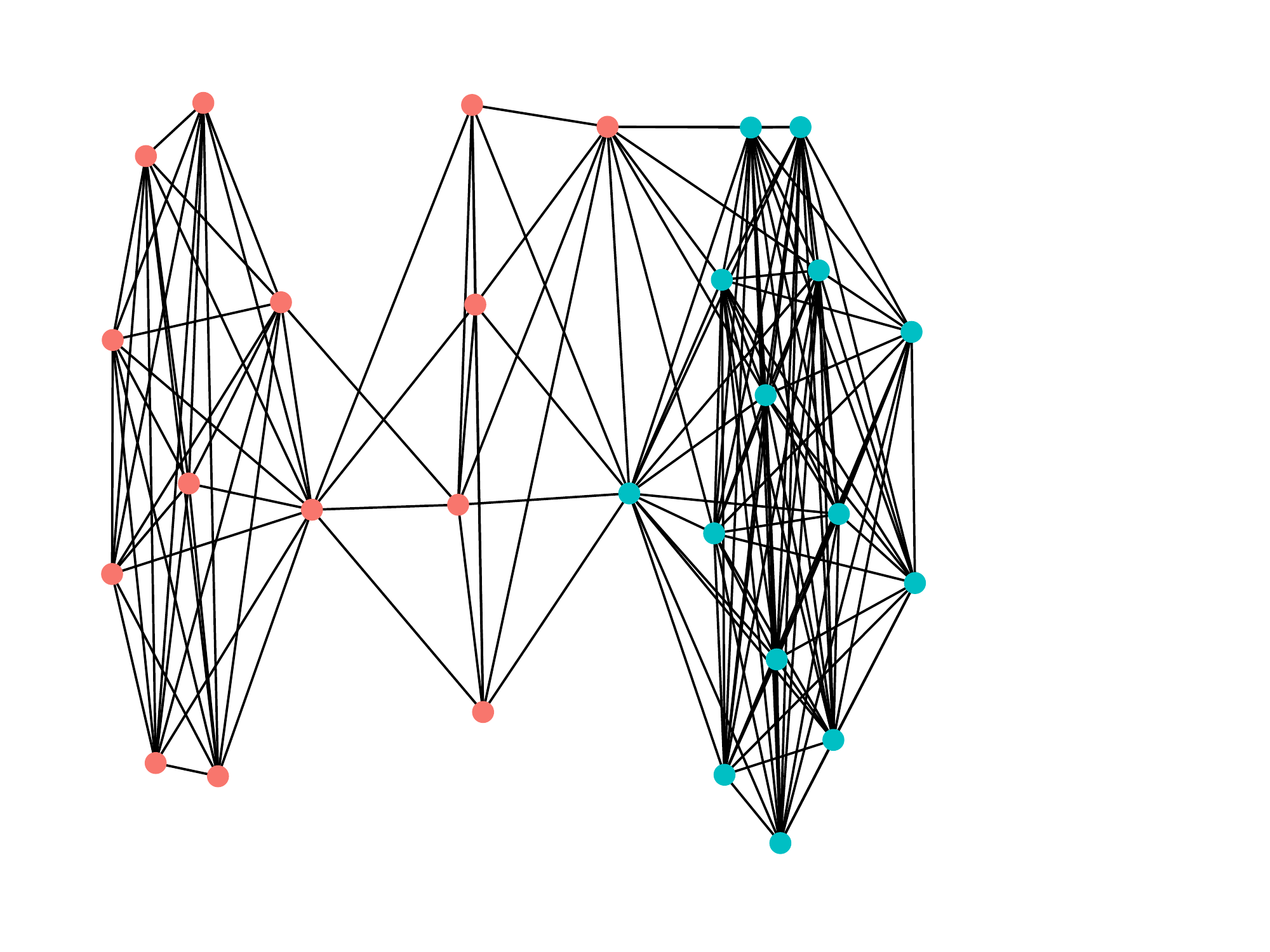}}}
    \hfill
    \subfloat[]{\scalebox{0.25}{\includegraphics{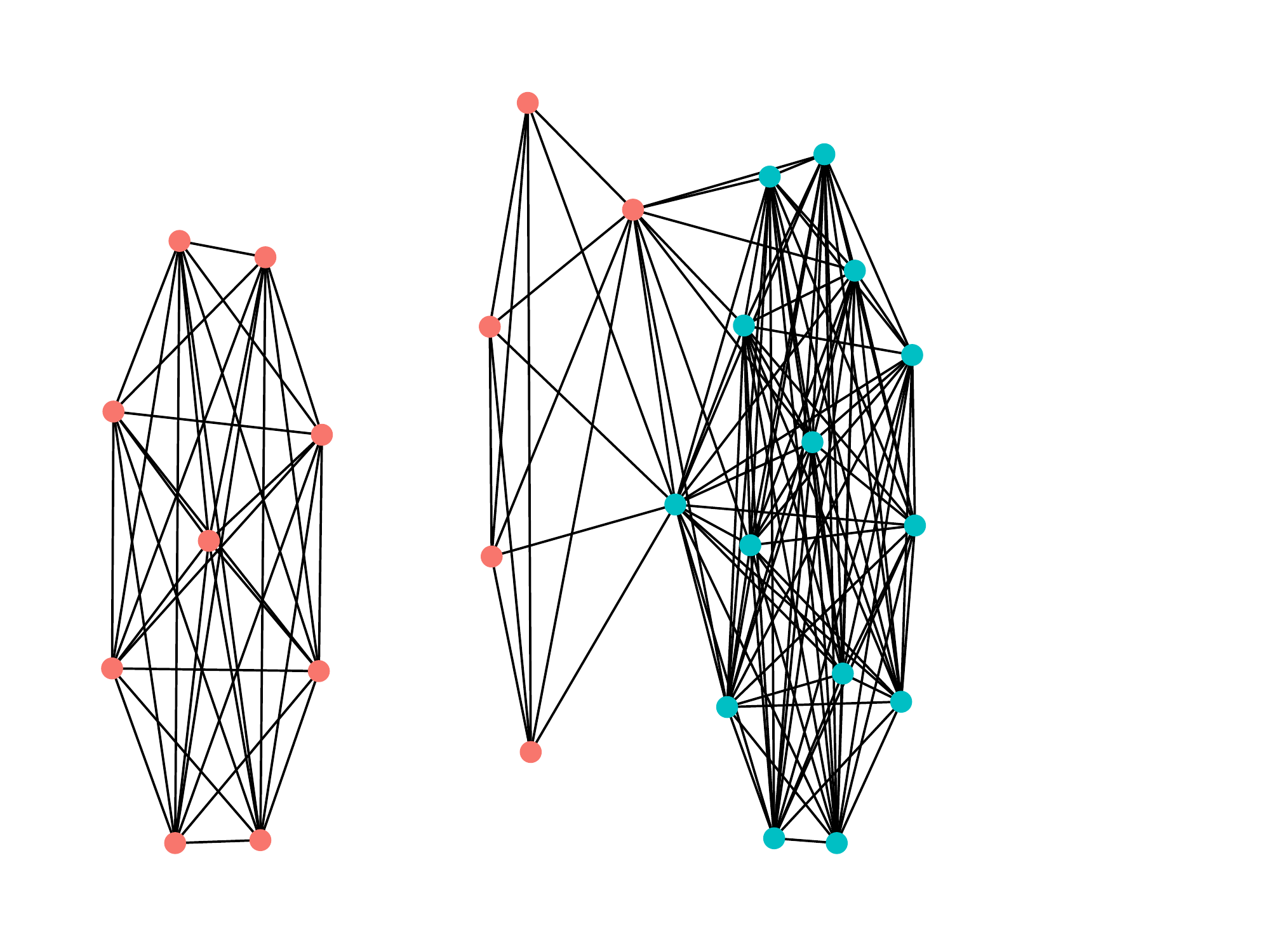}}}
    \hfill
    \subfloat[]{\scalebox{0.25}{\includegraphics{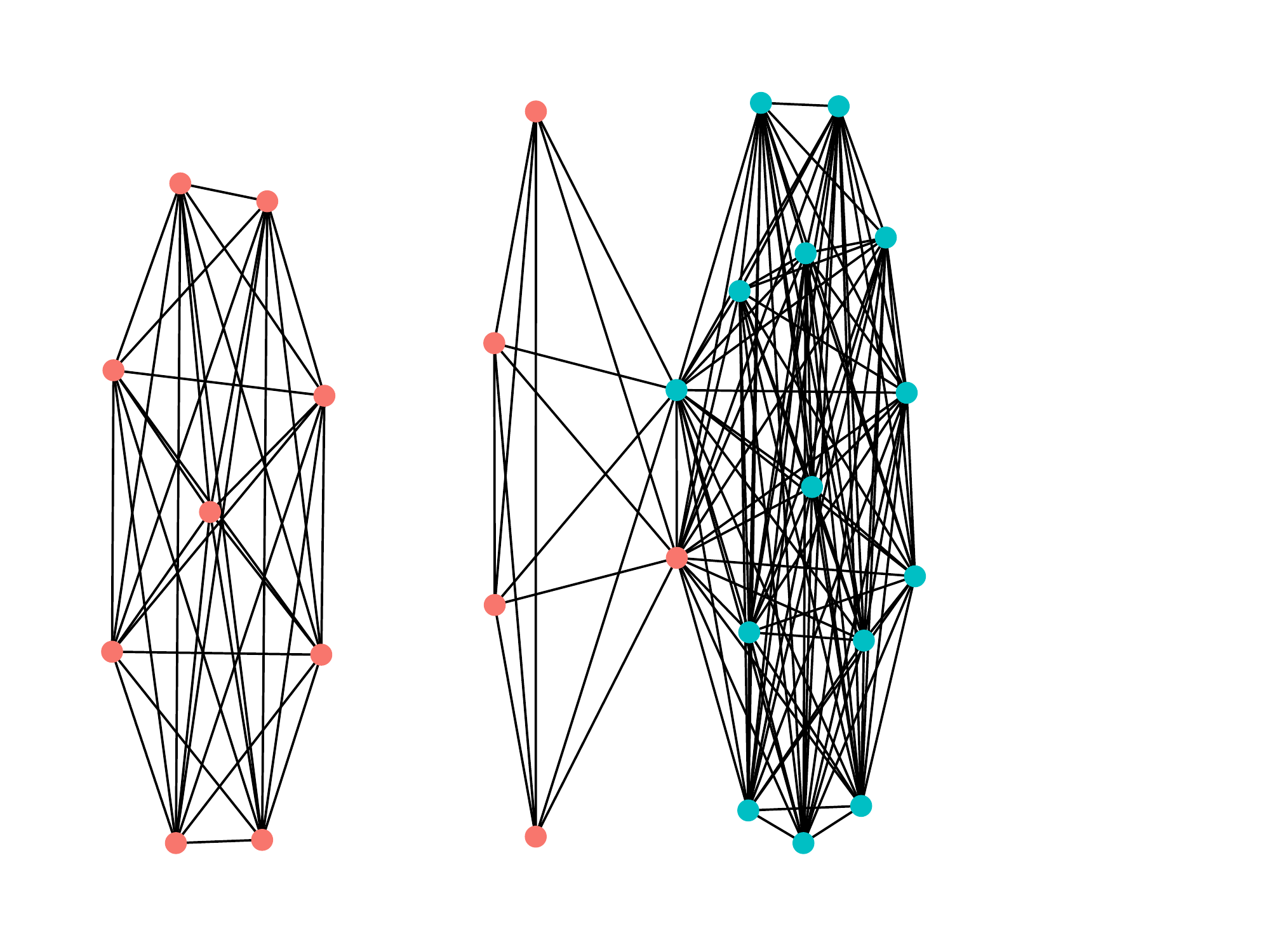}}}
    \caption{Simulated Network Analysis by Authenticity}
    \floatfoot{\footnotesize This figure plots the estimated network of simulated ``likes''. Panel (a) shows the case of politicians with no authenticity ($\gamma=0$); Panel (b) $\gamma=0.05$; Panel (c) $\gamma=0.1$; Panel (d) $\gamma=0.15$; Panel (e) $\gamma=0.2$}
      \label{fig:simulated_networks}
\end{figure}

\begin{figure}[H]
  \includegraphics[width=\linewidth]{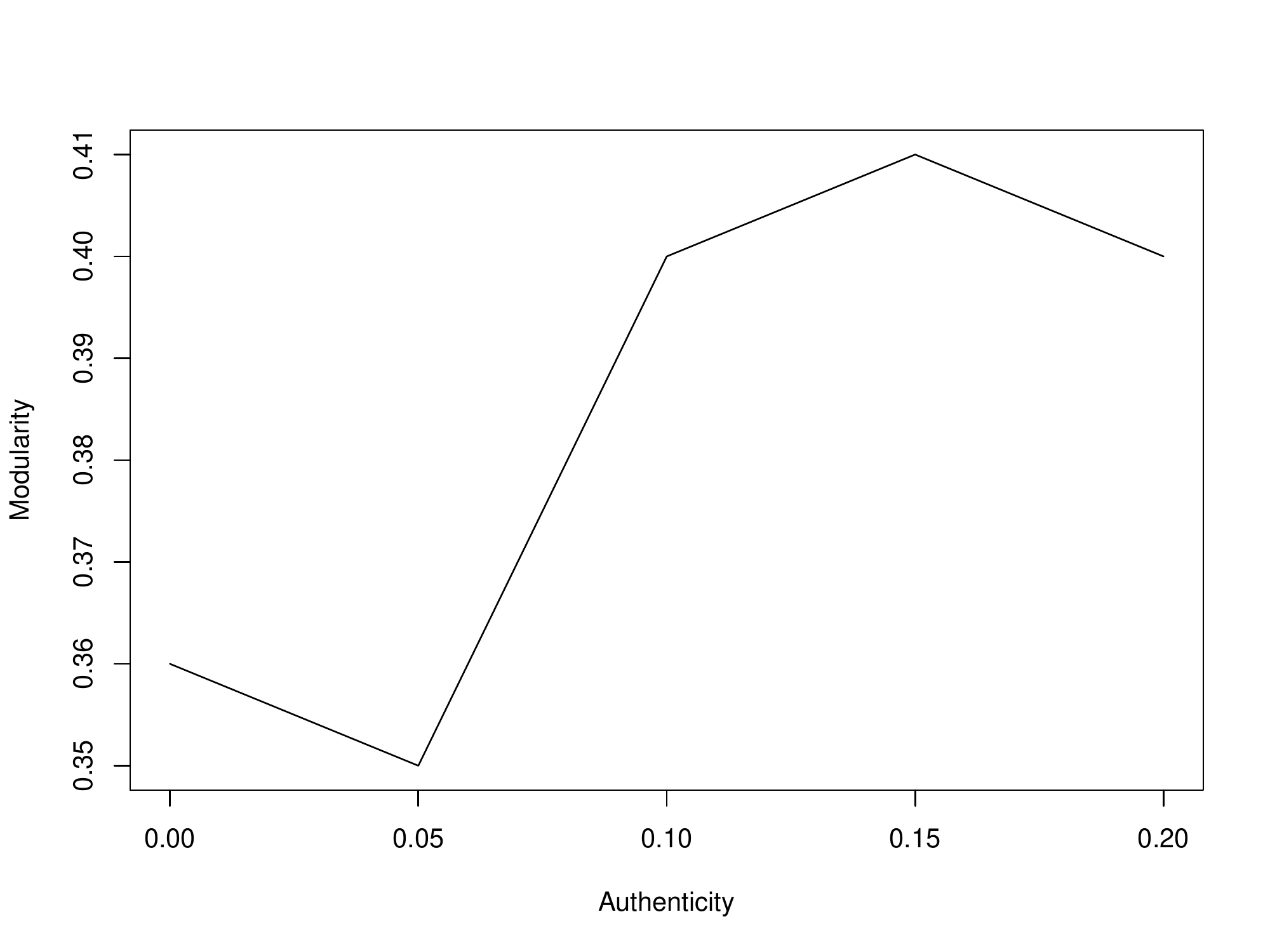}
  \caption{The figure plots the modularity of the simulated networks presented in Figure \ref{fig:simulated_networks} for different levels of authenticity ($\gamma$).}
  \label{fig:ModularityModel}
\end{figure}

In Figure \ref{fig:dist_elect_inc} we can see that our model generates a behavior that can explain how politicians at the extremes of the spectrum do not have incentives to move toward the center. Second, although not all politicians have incentives to interact with their political opponents, in the first panel of Figure \ref{fig:simulated_networks} we can see that our model still generates a fair amount of ``likes'' between political opponents. And this occurs in our self interest setting. In the following panels of Figure \ref{fig:simulated_networks}, we show how the simulated networks of ``likes'' changes for different levels of authenticity ($\gamma$), assuming homogeneous preferences. Visually, we can see that if all politicians are more authentic at the same time, the model generates ``echo chambers'' where only some politicians interact among each other. As we proposed in Section \ref{sec:networks_methodology}, in Figure \ref{fig:ModularityModel} we confirm that a systematic increase in politicians' authenticity is related to higher polarization, measured by networks' modularity.

Finally, we present a simulation based on heterogeneous levels of politicians' authenticity, which are randomly endowed independently of politicians' ideology. By assumption, authenticity is normally distributed with mean and standard deviation of 0.1, with a bound at zero. The simulated networks of ``likes'' is analyzed based on the following regression analysis:
\begin{equation}\label{eq:reg_model}
\text{Likes}_{i,j}=\alpha+\beta \cdot \text{Opponents}_{i,j}+\epsilon_{i,j}
\end{equation}
\noindent where $\text{Likes}_{i,j}$ measures the number of simulated ``likes'' between politician $i$ and $j$; $\text{Opponents}$ is a a dummy variable that takes a 1 if politicians $i$ and $j$ are opponents; $\epsilon_{i,j}$ is the residual.

The estimated coefficient ($\beta$) of the regression is presented in Table \ref{tab:reg_model}. As we can see, in our model opponents have a lower propensity to ``like'' each other's messages. In Figure X we present the relationship between the residual of Equation \ref{eq:reg_model} and the first principal component of politicians' ideology and authenticity. This relationship shows that ``likes'' between opponents are confined to politicians that have higher levels of authenticity and that are ideologically close. Importantly, this analysis suggest that, to the extent that we understand politicians' incentives, ``likes'' between opponents can be informative of politicians authenticity.
\newpage

\section{Empirical framework}\label{sec:emp_fram}
In Section \ref{sec:Model} we outline a public choice model to highlight the effect of political incentives and politicians' intrinsic authenticity on their propensity to ``like'' messages of other Congress members. The model generates the following main predictions. First, a politician that is less authentic is more  willing to ``like'' an opponent's message even if she disagrees with the ideological content of the message. Second, controlling for the electoral incentives, if a politician ``likes'' the message of an opponent, this observed behavior can be informative of politicians' authenticity and the ideological agreement between the two Congress members. Based upon our combination of voting data from Chilean Deputies and their corresponding interactions in Twitter, we empirically analyze the relationship between the interactions of the politician in Twitter and Congress.
Our analysis is conducted using a fixed effect regressions on the combined panel data that describes the historical interactions of politicians in Twitter and Congress. Our regression model is presented below:
\begin{equation}\label{eq:main_reg}
\begin{aligned}
Y_{i,j,t}=\alpha_i+\alpha_t+\beta\cdot\,\text{Likes}_{i,j,t}+\lambda\cdot\,\text{Likes}_{i,j,t}^2+\beta^{O}\cdot\,\text{Likes}_{i,j,t}\text{Opp}_{i,j}+\\\lambda^{O}\cdot\,\text{Likes}_{i,j,t}^2\cdot\text{Opp}_{i,j}+\theta_{i,j}+\epsilon_{i,j,t}
\end{aligned}
\end{equation}
where $Y_{i,j,t}$ is the numbers of ``likes'' or bills voted in favor between politician $i$ and $j$ during time window $t$, which are used interchangeable in the different specifications presented in Table \ref{tab:reg_model}; $\text{Likes}_{i,j,t}^2$ measures the number of ``likes'' between politician $i$ and $j$ squared, to capture potential nonlinear effects in the relationship of votes and ``likes''; $\text{Opp}_{i,j,t}$ is a dummy variable that takes a 1 if politician $i$ and $j$ are opponents;\footnote{The Opponent $\text{Opp}$ variable is constructed based on party affiliation.} $\alpha_i$ and $\alpha_t$ are politicians and time fixed effects, respectively; $\theta_{i,j}$ are two dummy variables that are used as non-time-varying controls, these are indicator functions that identify if politician $i$ follows politician $j$, and if politician $i$ and $j$ are opponents; $\epsilon_{i,j,t}$ is the residual.

Our main focus of analysis is related to the parameters $\lambda+\beta$ and $\lambda+\beta+\lambda^O$ of Equation \ref{eq:main_reg} because they measure how ``likes'' to political opponents is associated with the number of bills voted in favor by the two deputies. These estimates allow us to test our hypothesis of Twitter working as a public sphere. To the extent that these estimates are positive and significant, this would suggest that ``likes'' between political opponents are relevant signals of agreement.



\section{Results}\label{sec:results}
This section presents the results of a test that relates deputies' voting behavior to their propensity to ``like'' other politicians' messages in Twitter.

We begin with the study of deputies' behavior in Twitter. In columns (1) to (3) of Table \ref{tab:Regresiones} we document the coefficient estimates related to Equation \ref{eq:main_reg} using the number of ``likes'' as the dependent variable. Our findings suggest that deputies have a lower propensity to ``like'' their opponents messages, consistent with our theoretical model and the empirical evidence documented by \citet{halberstam2016homophily}. At the same time, our results also suggest that following links cannot explain ``likes'' when we control for political affiliation, which is consistent with the fact that politicians do want to read other congress members' ``tweets''.

Columns (4) to (6) of Table \ref{tab:Regresiones} use the number of bills voted in favor as the dependent variable. Interestingly, our results finds no relationship between bills voted in favor and coalition membership. This result is consistent with low levels of coalition unity during the analyzed period. Moreover, when we analyze columns (7) to (9) we can see that, during the studied period, the number of ``likes'' in Twitter is negatively related to the number of bills voted in favor. This finding suggest that a deputy could ``like'' the ``tweet'' of a coalition mate as a signal of group loyalty \citep{brennan1998expressive}, even when she opportunistically deviate from its coalition in Congress. Finally, in columns (8) and (9) we document our main finding. Giving a ``like'' to an opponent is positive and significantly correlated with future cross-cutting coalition in Congress. The differential effect between intra- and inter-coalition suggest that ``likes'' between political opponents work as costly signals of ideological agreement.
\begin{landscape}\label{tab:Regresiones}

\begin{table}[]
\centering
  \caption{Politicians interactions in Twitter and Congress}
\begin{tabular}{lccccccccc}
                                                         & (1)       & (2)      & (3)       & (4)     & (5)     & (6)     & (7)      & (8)       & (9)       \\
                                                         & Likes     & Likes    & Likes     & Votes   & Votes   & Votes   & Votes    & Votes     & Votes     \\
Opponents                                                & -0.812*** &          & -0.811*** & 0.275   &         & 0.763   &          & 0.612     & 0.486     \\
                                                         & (0.045)   &          & (0.043)   & (3.939) &         & (4.029) &          & (4.045)   & (4.026)   \\
Following                                                &           & 0.315*** & 0.003     &         & 1.461   & 1.754*  & 1.488    & 1.663*    & 1.735*    \\
                                                         &           & (0.041)  & (0.032)   &         & (1.375) & (0.686) & (1.368)  & (0.721)   & (0.695)   \\
Likes                                                    &           &          &           &         &         &         & -0.088** & -0.088*** & -0.345*** \\
                                                         &           &          &           &         &         &         & (0.022)  & (0.005)   & (0.068)   \\
Likes $\times$ \text{Opponent}        &           &          &           &         &         &         &          & 1.717*    & 2.957**   \\
                                                         &           &          &           &         &         &         &          & (0.797)   & (0.771)   \\
Likes$^2$                                                &           &          &           &         &         &         &          &           & 0.002**   \\
                                                         &           &          &           &         &         &         &          &           & (0.000)   \\
Likes$^2$ $\times$ \text{Opponent}           &          &           &         &         &         &          &           & -0.123    \\
                                                         &           &          &           &         &         &         &          &           & (0.102)   \\
Observations                                             & 48650     & 48650    & 48650     & 48650   & 48650   & 48650   & 48650    & 48650     & 48650     \\
Adjusted R-squared                                       & 0.011     & 0.005    & 0.011     & 0.091   & 0.092   & 0.092   & 0.092    & 0.092     & 0.092     \\
Fixed Effects                                            & \multicolumn{9}{c}{Time \& Deputy FE}                                                            
\end{tabular}
\end{table}
\end{landscape}


\section{Conclusion}\label{sec:con}
This paper tests the hypothesis that Twitter provides relevant information about politicians' legislative decisions. As opposed to other papers that have used politicians' Twitter data, we focus on the number of ``likes'' as an observable measure of agreement between Congress members.
Our empirical test intends to isolate the relationship between agreement in Twitter and Congress.
We find that ``likes'' are negative and significantly related to the number of bills voted in favor. We do, however, find a heterogeneous propensity to vote bills in favor if the ``like'' is given to an opponent. Our results can be rationalized by ``likes'' to coalition mates being signals of group loyalty instead of political agreement. On the other hand, ``likes'' to opponents would work as costly signals of ideological agreement as in our proposed political economy model. The implications of our results are related to voters ability to distinguish politicians' epistemic impartiality, an important condition for reaching political agreement as suggested by \citep{taylor2020agreeing}.

\begin{table}[htbp]
  \centering
  \caption{Analyzed time periods}
    \begin{tabular}{ccc}
    \multicolumn{2}{c}{\textbf{Votes}} &  \\
    \textbf{Start date} & \textbf{End date} & \textbf{Likes} \\
    April 2, 2019 & May 28, 2019 & May 27, 2019 \\
    September 3, 2019 & October 20, 2019 & October 28, 2019 \\
    December 30, 2019 & January 30, 2020 & February 21, 2020 \\
    May 12, 2020 & June 10, 2020 & June 11, 2020 \\
    November 3, 2020 & December 2, 2020 & December 2, 2020 \\
    \end{tabular}%
  \label{tab:DatesData}%
\end{table}%

\begin{table}[htbp]
  \centering
  \caption{Regression of Simulated Model}
    \begin{tabular}{ll}
          & Coefficients \\
    Opponents & -172.043*** \\
          & (10.614) \\
    Constant & 177.401*** \\
          & (10.467) \\
    Observations & 784 \\
    Adjusted R-squared & 0.251 \\
    \end{tabular}%
  \label{tab:reg_model}%
\end{table}%

\newpage

\appendix
\section{Network modularity}\label{sec:Community_detection}
A network cluster (communities) is a region of the network that is strongly connected within and relatively sparsely connected to the rest of the network. The common way of measuring how well a subdivision of a network into clusters capture the modular structure of the network is by network modularity \citep{Newman2006}. Mathematically, the network modularity $Q$ is given by:
\begin{equation}
    Q=\frac{1}{2m}\sum_{i,j}\Big(\text{a}_{i,j}-\frac{\text{k}_{i}\text{k}_{j}}{2m}\Big)\delta(\text{c}_{i},\text{c}_{j})
\end{equation}
where $\text{a}_{i,j}=1$ if nodes $i$ and $j$ are connected and $\text{a}_{i,j}=0$, $\text{otherwise}$. $m=\frac{1}{2}\sum_{i,j}\text{a}_{i,j}$ is the total number of edges in the network and $\text{k}_{i}=\sum_{j}a_{i,j}\delta(\text{c}_{i},\text{c}_{j})=1$ only if nodes $i$ and $j$ are in the same community. Otherwise, $\delta(c_{i},\text{c}_{j})=0$.
\section{Estimated Politicians Ideology}\label{sec:OLS_estimate}
\begin{figure}[H]
  \includegraphics[width=\linewidth]{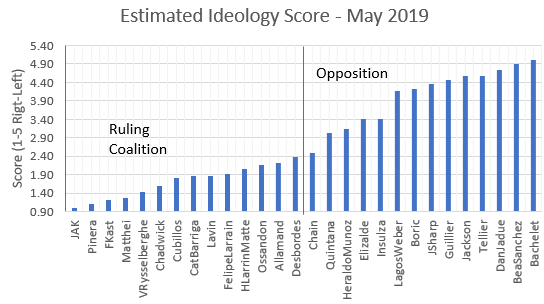}
  \caption{The figure plots the mean ideology of each politician in the model. Politicians are sorted from right to left.}
  \label{fig:Politician_Estimates2}
\end{figure}

In this section we propose a simple method for estimating politicians perceived ideology from public opinion surveys. The measure is constructed using a simple OLS regression:
\begin{equation}
\text{Opinion}_{j,k}\,=\alpha+\,\beta_{k}\, \text{Ideology}_{j}+\epsilon_j
\end{equation}
\noindent where $Opinion_{j}$ captures how favorable is the opinion of surveyed $j$ about politician $k$. Voters opinion is measured in a 1 (very bad) to 5 (very good) scale. $Ideology_{j}$ measures self reported ideology of surveyed $j$. Parameters that describe politicians ideology in the model ($\mu$ $\sigma$) are obtained from the mean and standard error of coefficients $\beta_{k}$. In the right and left extremes we have politicians with ideologies of 1 and 5, respectively. Ideology of the other politicians are re-adjusted using coefficients presented in Figure \ref{fig:Politician_Estimates1} in order to have a supply of politicians that go from 1 (right) to 5 (left). In Figure \ref{fig:Politician_Estimates2} we show the estimated levels of politicians from the CEP public opinion study as of May 2009.

\section{Median Voter Theorem}\label{sec:MedianVoter}
\begin{figure}[htbp]
  \includegraphics[width=\linewidth]{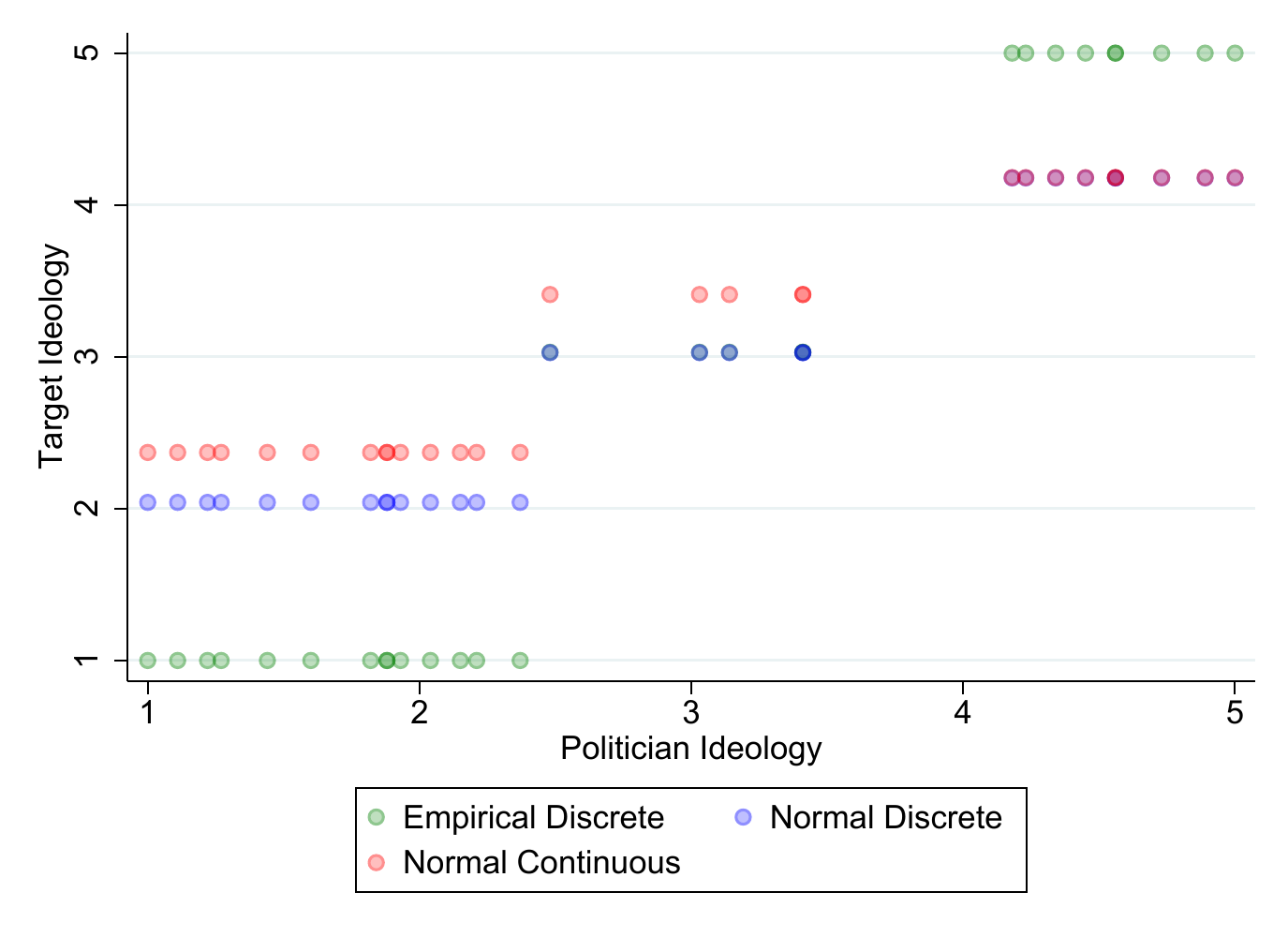}
  \caption{The figure shows how the electoral incentives are shaped by the distribution of voters. In the X-axis (Politician Ideology) is politicians' original ideology. In the Y-axis (Target Ideology), we have the ideology of the front-runner of the coalition for each politician. Empirical Discrete refers to the observed discrete distribution in the CEP opinion survey. Normal Discrete is a discrete distribution with the same mean and standard deviation of the voters' ideology. Normal Continuous corresponds to a continuous version of Normal Discrete.}
  \label{fig:dist_elect_inc}
\end{figure}

A standard benchmark in the public choice theory is the Median Voter Theorem of \citet{dow1998spatial}. In Figure \ref{fig:dist_elect_inc} we show that our calibration of voters introduces an incentive to some politicians to move toward the extremes. Contrarily, a normal distribution of voters' ideology (same mean and standard deviation) increases the incentives to move toward the center. This incentive is enhanced, if the population is more continuously distributed in the whole political spectrum.

\newpage

\bibliography{mybibfile}

\end{document}